\documentclass[11pt]{article}
\usepackage{amsfonts}
\usepackage{lscape}
\usepackage{latexsym,amsmath}
\usepackage{amssymb,array}
\makeatletter \oddsidemargin 0in \evensidemargin 0in \textwidth
16cm \RequirePackage[dvips]{graphicx} \textheight 20cm
\newtheorem{t1}{Theorem}[section]

\newtheorem{co}{Counterexample}[section]

\begin{document}
\title{Estimation of Reliability in the Two-Parameter Geometric Distribution}
\author{Sudhansu S.
Maiti\footnote{Corresponding author. E-mail Address:
dssm1@rediffmail.com},~~~~~Sudhir Murmu\\Department of Statistics, Visva-Bharati
University\\Santiniketan 731 235, India\\G. Chattopadhyay\\Department of Statistics, Calcutta University\\35, B. C. Road, Kolkata 700 019, India}
\date{}
\maketitle
\begin{abstract}
In this article, the reliabilities $R(t)=P(X\geq t)$, when $X$ follows two-parameter geometric distribution and $R=P(X\leq Y)$, arises under stress-strength setup, when X and Y assumed to follow two-parameter geometric independently have been found out. Maximum Likelihood Estimator (MLE) and an Unbiased Estimator (UE) of these have been derived. MLE and UE of the reliability of k-out-of-m system have also been derived. The estimators have been compared through simulation study.
\end{abstract}

\section{Introduction}
Various lifetime models have been proposed to represent lifetime data. Most of these models assume lifetime to be a continuous random variable. However, it is sometimes impossible or inconvenient to measure the life length of a device on a continuous scale. In practice, we come across situations where lifetimes are recorded on a discrete scale. Discrete life distributions have been mentioned by Barlow and Proschan [$\ref{bp}$]. Here one may consider lifetime to be the number of successful cycles or operations of a device before failure. For example, the bulb in xerox machine lights up each time a copy is taken. A spring may breakdown after completing a certain number of cycles of `to-and-fro' movements.\\
\hspace*{.2in} The study of discrete distributions in lifetime models is not very old. Yakub and Khan [$\ref{yk}$] considered the geometric distribution as a failure law in life testing and obtained various parametric and nonparametric estimation procedures for reliability characteristics. Bhattacharya and Kumar [$\ref{bk}$] discussed the parametric as well as Bayesian approach to the estimation of the mean life cycle and reliability for this model for complete as well as censored sample. Krishna and Jain [$\ref{kj}$] obtained classical and Bayes estimation of reliability for some basic system configurations. Modeling in terms of two-parameter geometric and estimation of its parameters and related functions are of special interest to a manufacturer who wishes to offer a minimum warranty life cycle of the items produced.\\
\hspace*{.2in} The two-parameter geometric distribution [abbreviated as $Geo (r, \theta)$] given by
\begin{equation}\label{eq1.1}
	P(X=x)=(1-\theta)\theta^{x-r}~~;~x=r, r+1, r+2, ...~~0<\theta<1,~ r\in\{0,1,2,...\},
\end{equation}
is the discrete analog of two-parameter exponential distribution. If $X$ follows a two-parameter exponential distribution, $\left[X\right]$, the integer part of $X$, has a two-parameter geometric distribution [see Kalbfleish and Prentice [$\ref{kp}$, Ch. 3]]. The reliability of a component when $X$ follows two-parameter geometric distribution is given by
\begin{equation}\label{eq1.2}
R(t)=\theta^{t-r};~~t=r, r+1, r+2, ....
\end{equation}
Laurent [$\ref{laurent}$] and Tate [$\ref{tate}$] obtained the uniformly minimum variance unbiased estimator (UMVUE) of the reliability function for the two-parameter exponential model. Different estimators of this reliability function have been discussed in Sinha [$\ref{sinha}$].\\
If a system consists of $m$ identical components each follows two-parameter geometric distribution, then the reliability of $k$-out-of-$m$ system is given by
\begin{equation}\label{eq1.3}
R_s(t)=P(X_{(m-k+1)}\ge t)=\sum_{i=k}^m {m \choose i}R(t)^i[1-R(t)]^{m-i};~~t=r, r+1, r+2, ...
\end{equation}
Special cases of $R_s(t)$ give series system (for $k=m$) and parallel system (for $k=1$).\\

\hspace*{.2in} In the stress-strength setup, $R=P(X\le Y)$ originated in the context of the reliability of a component of strength $Y$ subjected to a stress $X$. The component fails if at any time the applied stress is greater than its strength and there is no failure when $X \le Y$. Thus $R$ is a measure of the reliability of the component. Many authors considered the problem of estimation of $R$ in continuous setup in the past. Particularly, for the two-parameter exponential set up, Beg [$\ref{beg}$] derived the MLE and the UMVUE of $R$. In the discrete setup, the reference list is very limited. Maiti [$\ref{maiti95}$] has considered stress (or demand) $X$ and strength (or supply) $Y$ as independently distributed geometric random variables, whereas Sathe and Dixit [$\ref{sd}$] assumed as negative binomial variables, and derived both MLE and UMVUE of $R$. Maiti and Kanji [$\ref{mk}$] has derived some expressions of $R$ using a characterization and Maiti [$\ref{maiti05},\ref{maiti06}$] considered MLE, UMVUE and Bayes Estimation of $R$ for some discrete distributions useful in life testing. All the above mentioned works have been concentrated on one-parameter family of discrete distributions.\\
\hspace*{.2in} If $X$ and $Y$ follow two-parameter geometric distributions with parameters $(\theta_1, r_1)$ and $(\theta_2, r_2)$ respectively, then
\begin{eqnarray}\label{eq1.4}
R &=&\rho\theta_2^\delta~~~~~~~\mbox{for}~ \delta>0\nonumber\\&=&1-(1-\rho)\theta_1^{-\delta}~~\mbox{for}~\delta<0,
\end{eqnarray}where $\rho=\frac{1-\theta_1}{1-\theta_1\theta_2}$ and $\delta=r_1-r_2.$\\
Here we are interested to see whether similar estimators are obtained in case of the two-parameter geometric distribution, the discrete analog of the two-parameter exponential distribution. Then, it might be straightforward to use the two-parameter geometric distribution in the discrete life testing problem where a minimum warranty life cycle of the item is offered. In this article, we have found out some estimators of both $R(t)$ and $R_s(t)$ for complete as well as censored sample. Some estimators of $R$ have also been provided. The estimators have been compared through simulation study.\\
\hspace*{.2in}The paper is organized as follows. In section 2, we have derived MLE and UE of both $R(t)$ and $R_s(t)$. We have also derived MLE of these reliability functions for type-I censored sample. MLE and an unbiased estimator of $R$ have been found out in section 3. In section 4, simulation results have been reported. Section 5 concludes.
\section{Estimation of $R(t)$ and $R_s(t)$}
\begin{figure}[ht]
\centering
\includegraphics[width=13 cm,keepaspectratio]{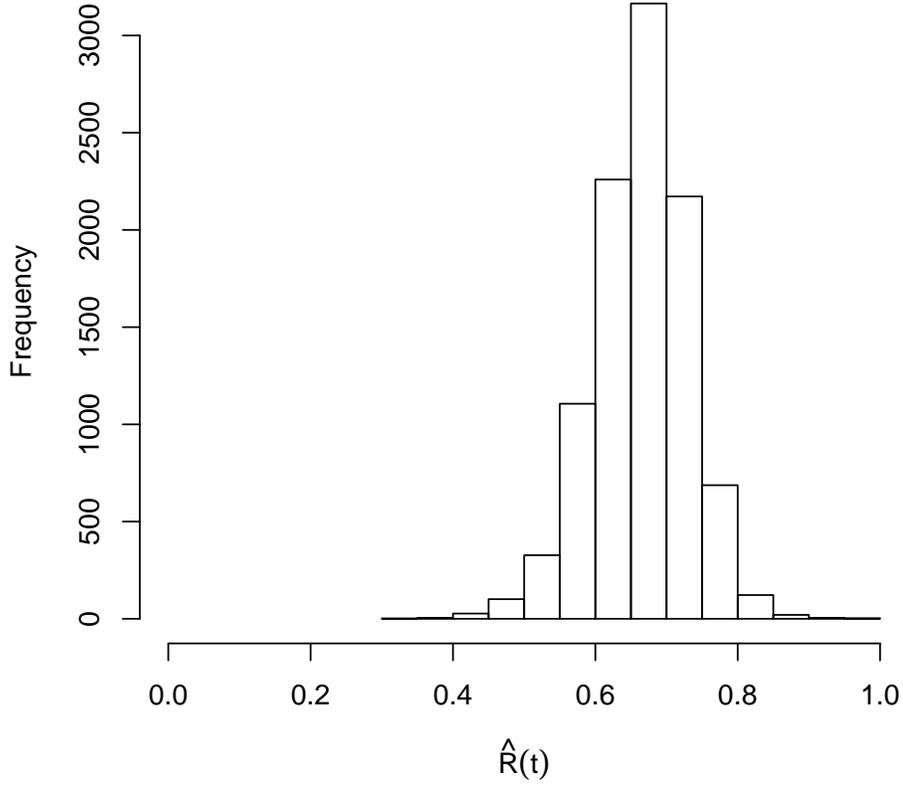}
\vspace{.1cm} \caption{Histogram of $\tilde{R}_U(t)$.}
\label{fig1}
\end{figure}

Let $(X_1, X_2, ..., X_n)$ be a random sample from $Geo(r, \theta)$ and $(X_{(1)}, X_{(2)}, ..., X_{(n)})$ be ordered sample. Maximum Likelihood Estimator of $r$ and $\theta$ are $X_{(1)}$ and $\frac{S}{n+S}$ respectively, where $S=\sum_{i=1}^n\left(X_i-X_{(1)}\right)=\sum_{i=1}^n\left(X_{(i)}-X_{(1)}\right).$ ML Estimators of $R(t)$ and $R_s(t)$ are given by
\begin{eqnarray}\label{eq2.1}
\hat{R}_M(t)&=&1~~~~~~~\mbox{for}~t\le X_{(1)} \nonumber\\&=&\left[\frac{S}{n+S}\right]^{t-X_{(1)}}~~\mbox{for}~t>X_{(1)}
\end{eqnarray}and
\begin{eqnarray}\label{eq2.2}
\hat{R}_{sM}(t)&=&1~~~~~~~\mbox{for}~t\le X_{(1)} \nonumber\\&=&\sum_{i=k}^m{m\choose i}\left[\hat{R}_M(t)\right]^i\left[1-\hat{R}_M(t)\right]^{m-i}~~\mbox{for}~t>X_{(1)}
\end{eqnarray}respectively.\\
\hspace*{.2in} Suppose we record the observations $(X_{(1)}, X_{(2)}, ..., X_{(p)}), p\le n$ that are failed before a pre-specified number of cycles c and remainings survive beyond c. Then, MLE of $r$ and $\theta$ are $X_{(1)}$ and $\frac{S^{*}}{p+S^{*}}$ respectively, where $S^{*}=\sum_{i=1}^p\left({X_{(i)}-X_{(1)}}\right)+(n-p)\{(c+1)-X_{(1)}\}.$ Hence  ML Estimators of $R(t)$ and $R_s(t)$ are given by
\begin{eqnarray}\label{eq2.3}
\hat{R^{*}}_M(t)&=&1~~~~~~~\mbox{for}~t\le X_{(1)} \nonumber\\&=&\left[\frac{S^{*}}{p+S^{*}}\right]^{t-X_{(1)}}~~\mbox{for}~t>X_{(1)}
\end{eqnarray}and
\begin{eqnarray}\label{eq2.4}
\hat{R^{*}}_{sM}(t)&=&1~~~~~~~\mbox{for}~t\le X_{(1)} \nonumber\\&=&\sum_{i=k}^m{m\choose i}\left[\hat{R^{*}}_M(t)\right]^i\left[1-\hat{R^{*}}_M(t)\right]^{m-i}~~\mbox{for}~t>X_{(1)}
\end{eqnarray}respectively.
\begin{t1}\label{th2.1}
$\left(X_{(1)}, S\right)$ is sufficient statistic for $\left(r, \theta\right)$. 
\end{t1}
Proof: Let $\underline X=(X_1, X_2, ..., X_n)$, we have to prove that $P(\underline X=\underline x|X_{(1)}=u,~S=s)$ does not depend on $r$ and $\theta$.\\Given $X_{(1)}=u$ and $S=s$, $\underline X$ is an $n$-dimensional random variable with domain $A_{u,s}=\{\underline x|x_{(1)}=u,~\sum_{i=2}^n x_i=s+(n-1)u\}$.\\For $\underline x\in A_{u,s}$,
$$P(\underline X=\underline x|X_{(1)}=u,~S=s)=\frac{P(\underline X=\underline x)}{\sum_{\underline y\in A_{u,s}}P(\underline X=\underline y)}$$and$$P(\underline X=\underline x)=\prod_{i=1}^n P(X_i=x_i)=(1-\theta)^n\theta^{s+n(u-r)}.$$Thus $P(\underline X=\underline x|X_{(1)}=u,~S=s)=\frac{1}{|A_{u,s}|}$, where ${|A_{u,s}|}$ is the number of elements in $A_{u,s}$. The number of elements in $A_{u,s}$ is the number of possible $n$-uplets $(x_1, x_2, ..., x_n)$ such that $x_{(1)}=u$ and $\sum_{i=2}^n x_i=s+(n-1)u$ which clearly does not depend on $\theta$ and $r$.\\
\hspace*{.2in} But $\left(X_{(1)}, S\right)$ is not complete as it is to be seen from the following counter example.
\begin{co}\label{counter1}
Let us define $g(.,~.)$ as
\begin{eqnarray*}
g\left(X_{(1)}, S\right)&=&1~~~\mbox{if}~X_{(1)}=r+2,~S=0\\&=&-1~~~\mbox{if}~X_{(1)}=r+1,~S=n\\&=&0~~~\mbox{otherwise}.
\end{eqnarray*}
Now $X_{(1)}$ and $S$ can take values $r+2$ and $0$ with the probability $(1-\theta)^n\theta^{2n}$ (for $X_{(2)}=r+2,~...,~X_{(n)}=r+2$), and $X_{(1)}$ and $S$ can take values $r+1$ and $n$ with the probability $(1-\theta)^n\theta^{2n}$ (one such particular situation is $X_{(2)}=r+2,~...,~X_{(n-1)}=r+2,~X_{(n)}=r+3$).\\
Therefore, it is found that $E_{r,~\theta}\left[g\left(X_{(1)}, S\right)\right]=0$ but $g\left(X_{(1)}, S\right)\ne0$.
\end{co}
The upcoming theorem will demonstrate the conditional distribution of $X$ for given $\left(X_{(1)}, S\right)$.
\begin{t1}\label{th2.2}
	The conditional distribution of $X$ given $\left(X_{(1)}, S\right)$ is as following:\\
	For $n=1$,$$P\left(X=x/X_{(1)}, S\right)=1$$
	For $n=2$, 
\begin{eqnarray*}
	P\left(X=x/X_{(1)}, S\right)&=&\frac{1}{2}~~~\mbox{if}~~x=X_{(1)}\\&=&\frac{1}{2}~~~\mbox{if}~~x=X_{(1)}+S
\end{eqnarray*}
For $n\ge3$, $S<n$,
\begin{eqnarray*}
	P\left(X=x/X_{(1)}, S\right)&=&{S-\left(x-X_{(1)}\right)+n-2\choose S-\left(x-X_{(1)}\right)}/{S+n-1\choose S}~~\mbox{if}~X_{(1)}\le x\le X_{(1)}+S\\&=&0~~~~~~~~~~~~~~~~~\mbox{otherwise.}
\end{eqnarray*}
For $n\ge3$, $S\ge n$,
\begin{eqnarray*}
	P\left(X=x/X_{(1)}, S\right)&=&{S+n-2\choose S}/\left\{{S+n-1\choose S}-{S-1\choose n-1}\right\}~~\mbox{if}~ x= X_{(1)}\\&=&\left\{{S-\left(x-X_{(1)}\right)+n-2\choose S-\left(x-X_{(1)}\right)}-{S-\left(x-X_{(1)}\right)-1\choose n-2}\right\}/\left\{{S+n-1\choose S}-{S-1\choose n-1}\right\}\\&&~~~~~~~~~~~~~~~~~~~~~~~~~~~~~~~~~~~~~~~~~~~~~~~~~~~\mbox{if}~X_{(1)}<x\le X_{(1)}+S-(n-1)\\&=&{S-\left(x-X_{(1)}\right)+n-2\choose S-\left(x-X_{(1)}\right)}/\left\{{S+n-1\choose S}-{S-1\choose n-1}\right\}\\&&~~~~~~~~~~~~~~~~~~~~~~~~~~~~~~~~~~~~~~~~~~~~~~~~~~~\mbox{if}~X_{(1)}+S-(n-1)<x\le X_{(1)}+S\\&=&0~~~~~~~~~~~~~~~~~\mbox{otherwise.}
\end{eqnarray*}
\end{t1}
Proof: Joint distribution of $X_1,~X_2,~...,~X_n$ is given by
$$P(X_1,~X_2,~...,~X_n)=(1-\theta)^n\theta^{\sum_{i=1}^n(X_i-r)}=(1-\theta)^n\theta^{S+n(X_{(1)}-r)},~~~r\le X_{(1)}.$$
Now, \begin{eqnarray*}
	P\left(X=x/X_{(1)}, S\right)&=&\frac{\sum_{X_2,~X_3,~...,~X_n}P(X=x,~X_2,~...,~X_n/X_{(1)},~S)}{\sum_{X_1,~X_2,~X_3,~...,~X_n}P(X_1,~X_2,~...,~X_n/X_{(1)},~S)}\\&=&\frac{\sum_{(X_2,~X_3,~...,~X_n/X_{(1)},~S)}1}{\sum_{(X_1,~X_2,~X_3,~...,~X_n/X_{(1)},~S)}1}.
\end{eqnarray*}
Here the denominator is equivalent to finding out the total number of ways in which $S$ indistinguishable balls can be placed in $n$ cells so that at least one cell remains empty. In general, if there are $r$ indistinguishable balls to be placed randomly in $k$ cells, then the number of distinguishable distributions is ${k+r-1\choose r}$ whereas the number of distinguishable distributions in which no cells remains empty is ${r-1\choose k-1}$. Therefore, we get the denominator as ${S+n-1\choose S}-{S-1\choose n-1}$ if $S\ge n$ and if $S<n$, it will be ${S+n-1\choose S}$. Similarly, the numerator is equivalent to finding out the total number of ways in which $S-\left(x-X_{(1)}\right)$ indistinguishable balls can be placed in $n-1$ cells so that at least one cell remains empty and hence, we get it as ${S-\left(x-X_{(1)}\right)+n-2 \choose S-\left(x-X_{(1)}\right)}-{S-\left(x-X_{(1)}\right)-1\choose n-2}$ if $S-\left(x-X_{(1)}\right)\ge n-1$ and if $S-\left(x-X_{(1)}\right)<n-1$, it will be ${S-\left(x-X_{(1)}\right)+n-2\choose S-\left(x-X_{(1)}\right)}$. Hence the theorem follows.\\
Since $\left(X_{(1)}, S\right)$ is sufficient but not complete statistic for $\left(r, \theta\right)$, we are handicapped of searching the UMVUE of any estimable function of these parameters using the Lehmann-Scheff$\acute{e}$ theorem. Hence, we will find an improved estimator of the reliability functions using the Rao-Blackwell theorem.\\
Define
\begin{eqnarray*}
Y&=&1~~~~~~~~~~~~~~~~\mbox{if}~~X_1\ge t\\&=&0~~~~~~~~~~~~~~~~\mbox{otherwise.}
\end{eqnarray*}
Then $R(t)=E(Y)=P\left(X_1\ge t\right)$. Using the Rao-Blackwell theorem, an unbiased estimator of $R(t)$ is given as follows:\\
for $n=1$,
\begin{eqnarray*}
\tilde{R}_U(t)&=&1~~~~~~~\mbox{if}~t\le X_{(1)}\\&=&0~~~~~\mbox{if}~~t>X_{(1)}.
\end{eqnarray*}
for $n=2$,
\begin{eqnarray*}
\tilde{R}_U(t)&=&1~~~~~~~\mbox{if}~t\le X_{(1)}\\&=&\frac{1}{2}~~~~~~~\mbox{if}~~X_{(1)}<t\le X_{(1)}+S\\&=&0~~~~~~~\mbox{if}~~t>X_{(1)}+S.
\end{eqnarray*}
for $n\ge 3$ and $S<n$,
\begin{eqnarray*}
\tilde{R}_U(t)&=&1~~~~~~~\mbox{if}~t\le X_{(1)} \\&=&\sum_{x=t}^{X_{(1)}+S}{S-\left(x-X_{(1)}\right)+n-2\choose S-\left(x-X_{(1)}\right)}/{S+n-1\choose S}~~\mbox{if}~X_{(1)}< t\le X_{(1)}+S\\&=&0~~~~~~~~~~~~~\mbox{if}~~t>X_{(1)}+S.
\end{eqnarray*}
It can also be written as
\begin{eqnarray*}
\tilde{R}_U(t)&=&1~~~~~~~\mbox{if}~t\le X_{(1)} \\&=&\sum_{x=t}^{X_{(1)}+S}\frac{n-1}{X_{(1)}+S+n-1}\prod_{j=1}^{n-2}\frac{\left(X_{(1)}+S+n-x-1-j\right)}{\left(X_{(1)}+S+n-1-j\right)}~\mbox{if}~X_{(1)}<t\le X_{(1)}+S\\&=&0~~~~~~~~~~~~~\mbox{if}~~t>X_{(1)}+S.
\end{eqnarray*}
For $n\ge3$, $S\ge n$,
\begin{eqnarray*}
	\tilde{R}_U(t)&=&1~~~~~~~~~~~~~~~~~~~~~~~~~~~~~~~~~~~~~~~~~~~~~~~~~~~\mbox{if}~t\le X_{(1)}\\&=&\sum_{x=t}^{X_{(1)}+S-(n-1)}\left\{{S-\left(x-X_{(1)}\right)+n-2\choose S-\left(x-X_{(1)}\right)}-{S-\left(x-X_{(1)}\right)-1\choose n-2}\right\}/\left\{{S+n-1\choose S}-{S-1\choose n-1}\right\}\\&&+\sum_{x=X_{(1)}+S-(n-1)+1}^{X_{(1)}+S}{S-\left(x-X_{(1)}\right)+n-2\choose S-\left(x-X_{(1)}\right)}/\left\{{S+n-1\choose S}-{S-1\choose n-1}\right\}\\&&~~~~~~~~~~~~~~~~~~~~~~~~~~~~~~~~~~~~~~~~~~~~~~~~~~~\mbox{if}~X_{(1)}<t\le X_{(1)}+S-(n-1)\\&=&\sum_{x=t}^{X_{(1)}+S}{S-\left(x-X_{(1)}\right)+n-2\choose S-\left(x-X_{(1)}\right)}/\left\{{S+n-1\choose S}-{S-1\choose n-1}\right\}\\&&~~~~~~~~~~~~~~~~~~~~~~~~~~~~~~~~~~~~~~~~~~~~~~~~~~~\mbox{if}~X_{(1)}+S-(n-1)<t\le X_{(1)}+S\\&=&0~~~~~~~~~~~~~~~~~\mbox{otherwise.}
\end{eqnarray*}
In other way the estimator $\tilde{R}_U(t)$ is to be UMVUE if it is uncorrelated with all unbiased estimator of zero. We take a class of unbiased estimator of zero as $U_0=\{u:\sum_{i=1}^n c_i X_i=u,~\sum_{i=1}^n c_i=1\}$. If $\tilde{R}_U(t)$ is UMVUE, then $Cov(U_0,~\tilde{R}_U(t))=0$ i.e. $Cov(1000.U_0,~1000.\tilde{R}_U(t))=0$. Analytical derivation seems to be intractable. We go for simulation study taking some particular choices of $(c_1,~c_2,~...,~c_n)$ and different $t$, and $1000$ covariances have been calculated and their averages have been shown in Tables 7-8. It is noticed that they are not uncorrelated and hence $\tilde{R}_U(t)$ is not UMVUE.\\
The variance of this unbiased estimator will be smaller than the unbiased estimator $\frac{\sum_{i=1}^nI(X_i\ge t)}{n}$, where $I(.)$ is the indicator function.\\
To study the asymptotic behavior of $\tilde{R}_U(t)$ we conduct a simulation study taking different values of parameters. $10000$ estimates of $\tilde{R}_U(t)$ and $\hat{R}_M(t)$, their variances, $95\%$ confidence limits and coverage probability (CP) have been shown in table 9. Histogram of $\tilde{R}_U(t)$ for $n=20,r=15,t=25,\theta=0.96$ has been shown in Figure $\ref{fig1}$.  In this set up the true reliability, $R(t)=0.6648326$. The figure is near normal. From the table 9, it is also evident from coverage probability point of view, $\hat{R}_M(t)$ is better if $0.02<R(t)<0.5$, otherwise $\tilde{R}_U(t)$ is better. From the table it is observed that asymptotic variance is approximately $\frac{R(t)(1-R(t))}{2n}$.\\
Define
\begin{eqnarray*}
Z&=&1~~~~~~~~~~~~~~~~\mbox{if at least}~k~\mbox{of}~~X_i\mbox{'s among}~X_1,~X_2,~...,~X_m~\mbox{are greater than or equal to}~t\\&=&0~~~~~~~~~~~~~~~~\mbox{otherwise.}
\end{eqnarray*}
Then $R_s(t)=E(Z)=\sum_{i=k}^m {m \choose i}[R(t)]^i[1-R(t)]^{m-i}$. Using the Rao-Blackwell theorem, an unbiased estimator of $R_s(t)$ is given by (for $2\le m<n$)
\begin{eqnarray*}
\tilde{R}_{sU}(t)&=&1~~~~~~~\mbox{if}~t\le X_{(1)} \\&=&\sum_{i=k}^m {m \choose i}[\tilde R_U(t)]^i[1-\tilde R_U(t)]^{m-i}~~\mbox{if}~X_{(1)}<t\le X_{(1)}+S\\&=&0~~~~~~~~~~~~~\mbox{if}~~t>X_{(1)}+S.
\end{eqnarray*}
\section{Estimation of $R$}
Let $(X_1, X_2, ..., X_{n_1})$ and $(Y_1, Y_2, ..., Y_{n_2})$ be random samples from $Geo(r_1, \theta_1)$ and $Geo(r_2, \theta_2)$ respectively. $\left(X_{(1)}, S_1\right)$ and $\left(Y_{(1)}, S_2\right)$ are defined in the same way as in section 2. Hence ML Estimator of $R$ is given by
\begin{eqnarray*}\hat{R}_M &=&\hat{\rho}\left(\frac{S_2}{n_2+S_2}\right)^{\hat{\delta}}~~~~~~~\mbox{for}~ \hat{\delta}>0\\&=&1-(1-\hat{\rho})\left(\frac{S_1}{n_1+S_1}\right)^{-\hat{\delta}}~~\mbox{for}~\hat{\delta}<0,
\end{eqnarray*}where $\hat{\rho}=\frac{n_1n_2+n_1S_2}{n_1n_2+n_1S_2+n_2S_1}$ and $\hat{\delta}=X_{(1)}-Y_{(1)}.$\\
\hspace*{.2in} We define censored scheme in the same way as in section 2, with pre-specified censored cycles $c_1$ and $c_2$ and with $p_1$ and $p_2$ censored observations.\\
Then, ML Estimator of $R$ is given by
\begin{eqnarray*}\hat{R^{*}}_M &=&\hat{\rho^{*}}\left(\frac{S^{*}_2}{p_2+S^{*}_2}\right)^{\hat{\delta}}~~~~~~~\mbox{for}~ \hat{\delta}>0\\&=&1-(1-\hat{\rho^{*}})\left(\frac{S^{*}_1}{p_1+S^{*}_1}\right)^{-\hat{\delta}}~~\mbox{for}~\hat{\delta}<0,
\end{eqnarray*}where $\hat{\rho^{*}}=\frac{p_1p_2+p_1S^{*}_2}{p_1p_2+p_1S^{*}_2+p_2S^{*}_1}$ and $\hat{\delta}=X_{(1)}-Y_{(1)}.$\\
Define
\begin{eqnarray*}
Z&=&1~~~~~~~~~~~~~~~~\mbox{if}~~X_1\le Y_1\\&=&0~~~~~~~~~~~~~~~~\mbox{otherwise.}
\end{eqnarray*}
Then $R=E(Z)=P\left(X_1\le Y_1\right)$. Application of the Rao-Blackwell theorem gives an unbiased estimator of $R$ as
\begin{eqnarray*}
\tilde{R}_U&=&\frac{1}{n_1}+\sum_{x=X_{(1)}+1}^{Y_{(1)}}f(x/X_{(1)},S_1)+\sum_{x=Y_{(1)}}^{min(W_1,W_2)}\sum_{y=x}^{W_2}f(x/X_{(1)},S_1)f(y/Y_{(1)},S_2)~~~\mbox{if}~X_{(1)}<Y_{(1)}\\&=&\frac{1}{n_1}+\sum_{x=X_{(1)}}^{min(W_1,W_2)}\sum_{y=x}^{W_2}f(x/X_{(1)},S_1)f(y/Y_{(1)},S_2)~~~\mbox{if}~X_{(1)}=Y_{(1)}\\&=&\frac{1}{n_1}\sum_{y=X_{(1)}}^{W_2}f(y/Y_{(1)},S_2)+\sum_{x=X_{(1)}+1}^{min(W_1,W_2)}\sum_{y=x}^{W_2}f(x/X_{(1)},S_1)f(y/Y_{(1)},S_2)~~~\mbox{if}~X_{(1)}>Y_{(1)}
\end{eqnarray*} where, $W_1=X_{(1)}+S_1$, $W_2=Y_{(1)}+S_2$. The variance of this unbiased estimator will be smaller than the unbiased estimator $\frac{1}{n_1n_2}\sum_{i=1}^{n_1}\sum_{j=1}^{n_2}I(X_i<Y_j)$.
\section{Simulation Study}
\subsection{Discussion on simulation results relating to $R(t)$ and $R_s(t).$}
In order to have an idea about the selection of an estimator between Maximum Likelihood Estimator (MLE) and Unbiased Estimator (UE), Mean Squared Errors (MSEs) and hence percent relative efficiency using these MSEs have been calculated for $R(t)$ and $R_s(t)$. We generate sample of size $n$ and on the basis of this sample, calculate MLE and UE. MLEs have been calculated for complete as well as censored (type-I defined in earlier section) samples. 10000 such estimates have been calculated and results, on the basis of these estimates have been reported in Tables $1-6$. Initial set up for parameters has been chosen as $(n=20,~r=15,~c=25,~\theta=0.8,~t=25,~k=2,~m=8)$. Each table has been prepared considering different choices of a particular parameter, keeping others fixed at initial set up. All simulations and calculations have been done using R-Software and algorithms used can be obtained by contacting the corresponding author. Different columns of a table are as follows:\\
\hspace*{.2in}Col.1: Component Reliability, $R(t)$\\
\hspace*{.2in}Col.2: Average of MLEs of $R(t)$ for complete sample\\
\hspace*{.2in}Col.3: Average of MLEs of $R(t)$ for censored sample\\
\hspace*{.2in}Col.4: Percent relative efficiency of MLE of $R(t)$ for complete sample to that of censored\\\hspace*{.6in}   sample\\
\hspace*{.2in}Col.5: Average of UEs of $R(t)$\\
\hspace*{.2in}Col.6: Percent relative efficiency of UE to MLE of $R(t)$\\
\hspace*{.2in}Col.7: System Reliability, $R_s(t)$\\
\hspace*{.2in}Col.8: Average of MLEs of $R_s(t)$ for complete sample\\
\hspace*{.2in}Col.9: Average of MLEs of $R_s(t)$ for censored sample\\
\hspace*{.2in}Col.10: Percent relative efficiency of MLE of $R_s(t)$ for complete sample to that of censored\\\hspace*{.7in}sample\\
\hspace*{.2in}Col.11: Average of UEs of $R_s(t)$\\
\hspace*{.2in}Col.12: Percent relative efficiency of UE to MLE of $R_s(t)$\\
\hspace*{.2in}For estimation of component as well as system reliability, MLE performs quite well from efficiency point of view except for very less reliable component whose importance in practice is not so much meaningful (Table 1 and Table 2, Col.6 and Col.12). Even though the MLE is biased, the combined effect of variance and bias is less than the variance of the proposed UE. If sample size $n$ increases, as expected, MLE and UE for component as well as system reliability become equally efficient (Table $4$, Col.6 and Col.12). If censored number of cycle $c$ increases, MLEs for complete as well as censored sample become equally efficient (Table 3, Col.4 and Col.10).
\subsection{Discussion on simulation results relating to $R$.}
In order to have an idea about the nature of the MLEs under complete sample and censored sample, we have calculated percent relative efficiency of $\hat{R}_M$ with respect to $\hat{R^{*}}_M$ i.e.$\frac{\mbox{MSE of}~\hat{R^{*}}_M\times 100}{\mbox{MSE of}~\hat{R}_M}$ and presented in tables for $r_1=5,~10$, $r_2=10,~5$, $\theta_1=0.7,~0.8$, $\theta_2=0.7,~0.8$ and for different values of $c_1$ and $c_2$ (Table 10-13). Here, we take  $n_1=n_2=10$, and $1000$ estimates of $R$ have been taken for calculating MSEs. We observe that, as expected, MLE of $R$ for censored sample approaches to that of complete sample as both $c_1$ and $c_2$ increase. Fixing any one of $c_1$ and $c_2$, and increasing the remaining, efficiency of MLE for censored sample sometimes increases but there is no guarantee.\\
\hspace*{0.2in} In order to have an idea about the selection of an estimator between UE and MLE, we have calculated MSEs of Estimates of $R$. We prepared tables (Table14-18) for 
(i) $\theta_1=0.1,~\theta_2=0.1$,
(ii) $\theta_1=0.5,~\theta_2=0.5$,
(iii) $\theta_1=0.8,~\theta_2=0.2$,
(iv) $\theta_1=0.9,~\theta_2=0.9$,
(v) $\theta_1=0.2,~\theta_2=0.8$,and
$r_1$ and $r_2$ equal to 5, 10, 15 and 20, $n_1=n_2=10.$  In these tables, values in 1st row indicates true value of $R$, 2nd and 3rd rows indicate average of estimate of MLE and UE of $R$, and 4th and 5th rows indicate MSEs of MLE and UE of $R$ respectively.\\
\hspace*{0.2in} We observe that in almost all cases, MLE is better that UE for $R$ in mean square error sense. Therefore, as soon as we entered to unbiased class, we are losing some efficiency. It is to be noticed that MLE in this case is not an unbiased estimator. Moreover, MLE has a computational ease.
\section{Concluding Remark}
\hspace*{.2in}This paper takes into account the inferential aspects
of reliability with the two-parameter geometric lifetime. The continuous
distributions are widely referenced probability laws used in
reliability and life testing for continuous data. When the lives
of some equipments and components are being measured by the number
of completed cycles of operations or strokes, or in case of
periodic monitoring of continuous data, the discrete distribution
is a natural choice. At the same time, if a minimum warranty life cycle of the items are provided, the two-parameter geometric
distribution is the simplest but an important choice. Under this set up estimators of reliability functions- under mission time as well as under stress-strength
set up, have been viewed. It is interesting to note that, unlike the case
of the two-parameter exponential, the estimators of the parameters of the two-parameter geometric distribution are not complete. As a result, we only get unbiased estimators of the reliability functions for the two-parameter geometric set up.\\In most of the situations, MLE gives better result than the UE in mean square sense. As soon as we entered to unbiased class, we are loosing some efficiency. If one is ready to sacrifice the unbiased criteria of the estimator, the MLE in this case is preferable. It is to be noticed here that MLE is not an unbiased estimator. Moreover, MLE has a computational ease.

\pagebreak
\begin{landscape}
\begin{center}{\bf Table 1: Calculations relating to $R(t)$ and $R_s(t)$}\\$~n=20,~r=15,~c=25,~\theta=0.8,~k=2,~m=8$\\
\vspace{0.2in}
{\small\begin{tabular}{|c|c|c|c|c|c|c|c|c|c|c|c|c|}
\hline{$t$}&Col.1&Col.2&Col.3&Col.4&Col.5&Col.6&Col.7&Col.8&Col.9&Col.10&Col.11&Col.12\\
\hline16&0.8&0.79393&0.79491&104.89&0.80022&113.64&0.99992&0.999708&0.999706&100.15&0.99977&138.10\\
\hline17&0.64&0.62917&0.63338&106.57&0.63867&119.58&0.99571&0.982063&0.982062&100.31&0.98545&128.81\\
\hline18&0.512&0.49571&0.50086&106.73&0.51097&109.54&0.96979&0.928259&0.929258&99.76&0.93871&120.28\\
\hline19&0.4096&0.39881&0.40459&110.55&0.41452&98.88&0.90330&0.836803&0.838851&100.28&0.85352&111.01\\
\hline20&0.32768&0.31945&0.32699&113.34&0.33407&92.92&0.79549&0.722729&0.729116&101.38&0.74399&102.54\\
\hline25&0.10737&0.10622&0.11228&131.55&0.10647&80.50&0.20909&0.220599&0.236196&125.40&0.22473&83.46\\
\hline30&0.03518&0.04018&0.04422&151.20&0.03661&83.31&0.03009&0.056895&0.067862&168.06&0.05466&83.63\\
\hline31&0.02814&0.03461&0.03810&151.25&0.03011&89.96&0.01981&0.044496&0.054024&178.83&0.04043&91.94\\
\hline35&0.01153&0.01476&0.01754&212.54&0.01108&104.80&0.00355&0.011759&0.017823&334.81&0.00923&101.20\\
\hline40&0.00378&0.00692&0.00868&210.03&0.00473&122.46&0.00039&0.003788&0.006659&375.59&0.00271&107.04\\
\hline45&0.00123&0.00271&0.00342&226.31&0.00157&174.54&$4.26\times10^{-5}$&0.000858&0.001667&737.01&0.00048&189.46\\
\hline
\end{tabular}}
\end{center}
\pagebreak
\begin{center}{\bf Table 2: Calculations relating to $R(t)$ and $R_s(t)$}\\$~n=20,~r=15,~c=25,~\theta=0.8,~t=25,~m=8$\\
\vspace{0.2in}
\begin{tabular}{|c|c|c|c|c|c|c|c|c|c|c|c|c|}
\hline{$k$}&Col.1&Col.2&Col.3&Col.4&Col.5&Col.6&Col.7&Col.8&Col.9&Col.10&Col.11&Col.12\\
\hline1&0.10737&0.10841&0.11121&116.64&0.11827&78.010&0.59695&0.56534&0.57109&105.37&0.59365&94.259\\
\hline3&0.10737&0.10821&0.11118&118.38&0.11805&78.110&0.45796&0.06392&0.06981&137.86&0.07894&63.170\\
\hline6&0.10737&0.10829&0.11124&119.96&0.11813&78.048&$3.53\times10^{-05}$&0.000263&0.000408&374.98&0.000436&39.115\\
\hline8&0.10737&0.10933&0.11229&118.16&0.11929&77.648&$1.76\times10^{-08}$&$7.12\times10^{-07}$&$1.51\times10^{-06}$&1081.50&$1.46\times10^{-06}$&25.299\\
\hline
\end{tabular}
\end{center}
\begin{center}{\bf Table 3: Calculations relating to $R(t)$ and $R_s(t)$}\\$~n=20,~r=15,~\theta=0.8,~t=25,~k=2,~m=8$\\
\vspace{0.2in}
{\footnotesize\begin{tabular}{|c|c|c|c|c|c|c|c|c|c|c|c|c|}
\hline{$c$}&Col.1&Col.2&Col.3&Col.4&Col.5&Col.6&Col.7&Col.8&Col.9&Col.10&Col.11&Col.12\\
\hline20&0.10737&0.10790&0.11599&164.86&0.11772&78.104&0.20909&0.21959&0.24290&160.05&0.24860&76.145\\
\hline25&0.10737&0.10819&0.11123&118.83&0.11801&78.133&0.20909&0.22043&0.22924&118.50&0.24941&76.362\\
\hline30&0.10737&0.10829&0.10946&107.18&0.11814&78.029&0.20909&0.22059&0.22397&107.18&0.24964&76.268\\
\hline35&0.10737&0.10899&0.10962&103.95&0.11892&77.747&0.20909&0.22251&0.22432&103.93&0.25179&76.101\\
\hline40&0.10737&0.10842&0.10866&101.60&0.11828&78.084&0.20909&0.22093&0.22160&101.54&0.24995&76.446\\
\hline45&0.10737&0.10772&0.10780&100.63&0.11751&78.329&0.20909&0.21914&0.21935&100.58&0.24801&76.463\\
\hline
\end{tabular}}
\end{center}
\pagebreak
\begin{center}{\bf Table 4: Calculations relating to $R(t)$ and $R_s(t)$}\\$~r=15,~c=25,~\theta=0.8,~t=25,~k=2,~m=8$\\
\vspace{0.2in}
{\footnotesize\begin{tabular}{|c|c|c|c|c|c|c|c|c|c|c|c|c|}
\hline{$n$}&Col.1&Col.2&Col.3&Col.4&Col.5&Col.6&Col.7&Col.8&Col.9&Col.10&Col.11&Col.12\\
\hline10&0.10737&0.10599&0.11160&127.07&0.12610&62.804&0.20909&0.21957&0.23384&122.73&0.27625&63.291\\
\hline15&0.10737&0.10769&0.11140&120.18&0.12087&72.608&0.20909&0.22115&0.23150&118.81&0.25938&71.594\\
\hline20&0.10737&0.10909&0.11298&118.70&0.11984&77.429&0.20909&0.22491&0.23418&118.39&0.25443&75.910\\
\hline25&0.10737&0.10819&0.11075&117.64&0.11599&81.963&0.20909&0.21916&0.22675&118.07&0.24232&79.985\\
\hline50&0.10737&0.10830&0.10965&114.09&0.11215&90.156&0.20909&0.21598&0.22022&114.83&0.22757&88.475\\
\hline100&0.10737&0.10761&0.10823&111.36&0.10949&95.064&0.20909&0.21199&0.21403&111.92&0.21772&93.938\\
\hline200&0.10737&0.10735&0.10769&110.46&0.10829&97.569&0.20909&0.21011&0.21124&110.76&0.21297&96.925\\
\hline
\end{tabular}}
\end{center}
\begin{center}{\bf Table 5: Calculations relating to $R(t)$ and $R_s(t)$}\\$~n=20,~c=25,~\theta=0.8,~t=25,~k=2,~m=8$\\
{\footnotesize\begin{tabular}{|c|c|c|c|c|c|c|c|c|c|c|c|c|}
\hline{$r$}&Col.1&Col.2&Col.3&Col.4&Col.5&Col.6&Col.7&Col.8&Col.9&Col.10&Col.11&Col.12\\
\hline0&0.00377&0.00552&0.00556&101.87&0.00471&101.81&0.00039&0.00182&0.00185&94.865&0.00162&88.625\\
\hline5&0.01152&0.01420&0.01436&106.003&0.01366&86.929&0.00355&0.00885&0.00914&113.05&0.00904&79.946\\
\hline10&0.03518&0.03885&0.03976&112.85&0.04073&78.387&0.03009&0.04675&0.04916&120.39&0.05214&72.434\\
\hline15&0.10737&0.10779&0.11066&119.41&0.11756&78.399&0.20909&0.21937&0.22761&118.72&0.24815&76.659\\
\hline20&0.32768&0.31915&0.32740&138.77&0.34411&89.196&0.79548&0.75086&0.75481&117.19&0.79071&121.971\\
\hline
\end{tabular}}
\end{center}
\pagebreak
\begin{center}{\bf Table 6: Calculations relating to $R(t)$ and $R_s(t)$}\\$~n=20,~r=15,~c=25,~t=25,~k=2,~m=8$.\\
{\footnotesize\begin{tabular}{|c|c|c|c|c|c|c|c|c|c|c|c|c|}
\hline{$\theta$}&Col.1&Col.2&Col.3&Col.4&Col.5&Col.6&Col.7&Col.8&Col.9&Col.10&Col.11&Col.12\\
\hline0.4&0.00010&0.00032&0.00032&101.01&0.00012&337.78&$3.07\times10^{-7}$&$1.49\times10^{-5}$&$1.50\times10^{-5}$&107.31&$4.39\times10^{-6}$&491.22\\
\hline0.5&0.00097&0.00183&0.00184&103.22&0.00099&204.17&$2.65\times10^{-5}$&0.00028&0.00029&120.32&0.00013&230.03\\
\hline0.6&0.00605&0.00823&0.00831&104.67&0.00597&136.80&0.00099&0.00367&0.00367&111.03&0.00236&162.16\\
\hline0.7&0.02824&0.03160&0.03213&110.33&0.02798&102.49&0.01995&0.03348&0.03482&118.42&0.02875&112.80\\
\hline0.8&0.10737&0.10815&0.11123&118.94&0.10703&90.47&0.20909&0.22019&0.22923&119.02&0.21801&92.17\\
\hline0.9&0.34867&0.34157&0.34866&141.31&0.34973&97.45&0.82891&0.78709&0.78698&120.64&0.79895&105.99\\
\hline0.93&0.48398&0.47655&0.48212&165.12&0.48399&105.66&0.95725&0.93552&0.92884&155.35&0.94084&117.34\\
\hline0.96&0.66483&0.66788&0.66198&198.66&0.66456&118.19&0.99731&0.99469&0.99086&355.23&0.99487&120.60\\
\hline0.99&0.90438&0.93525&0.90791&173.29&0.90454&119.21&0.99999&0.99999&0.99996&390.01&0.99999&121.53\\
\hline
\end{tabular}}
\end{center}
\pagebreak
\begin{center}{\bf Table 7: Unbiased Estimator}\\$n=10,~r=15,~c=25,~\theta=0.8,~k=2,~m=8$\\
\vspace{0.2in}
{\footnotesize\begin{tabular}{|c|c|c|c|c|c|}
\hline{$Combination~|~~t$} &20&25&30&35&40\\
\hline
$+1-1+1-1+1-1+1-1+1-1$&$334142.2$&$30824.96$&$8892.633$&$13735.52$&$-9741.06$\\\hline
$+1+1+1+1+1-1-1-1-1-1$&$37026.8$&$2746.153$&$87121.23$&$-23797.94$&$4468.285$\\\hline
$+1+1~0~0~0~0~0~0-1-1$&$-78456.49$&$-31337.49$&$37854.66$&$843.4048$&$-17.16015$\\\hline
$+1~0~0~0~0~0~0~0~0-1$&$50794.91$&$-36774.63$&$14240.32$&$1158.73$&$-2478.892$\\\hline
Reliability&$0.32768$&$0.10737$&$0.03518$&$0.01152$&$0.00378$\\
\hline
\end{tabular}}
\end{center}
\begin{center}{\bf Table 8: Unbiased Estimator}\\$n=10,~r=15,~c=25,~t=25,~k=2,~m=8$\\
\vspace{0.2in}
{\footnotesize\begin{tabular}{|c|c|c|c|c|c|c|c|c|c|}
\hline{$Combination\downarrow~|~~\theta\rightarrow$} &0.4&0.5&0.6&0.7&0.8&0.9&0.93&0.96&0.99\\
\hline
$+1-1+1-1+1-1+1-1+1-1$&$277.34$&$1297.87$&$-7990.63$&$-25294.98$&$29730.69$&$466268.7$&$-626848.4$&$1314555$&$2910383$\\\hline
$+1+1+1+1+1-1-1-1-1-1$&$128.73$&$603.68$&$2339.78$&$2376.59$&$-12487.09$&$-378417.4$&$-551749.3$&$5398500$&$-532512.9$\\\hline
$+1+1~0~0~0~0~0~0-1-1$&$78.22$&$-118.91$&$3483.52$&$21081.09$&$-60072.55$&$-107717.3$&$-605752.3$&$1757503$&$-13004957$\\\hline
$+1~0~0~0~0~0~0~0~0-1$&$132.79$&$34.93$&$-941.95$&$3206.99$&$-5265.09$&$120432.4$&$-695584.4$&$530686.8$&$-9375230$\\\hline
Reliability&$0.000105$&$0.000977$&$0.006047$&$0.028248$&$0.107374$&$0.348678$&$0.483982$&$0.664833$&$0.904382$\\
\hline
\end{tabular}}
\end{center}
\pagebreak
\begin{center}{\bf Table 9: Calculation of Confidence Interval and CP of $R(t)$}\\$n=20,~r=15,~c=25,~t=25,~Repetation=10000$\\
\vspace{0.2in}
{\footnotesize\begin{tabular}{|l|l|l|l|l|l|l|l|l|l|}
\hline{$\theta$}&Reliability &\multicolumn{4}{|c|}{Unbiased Estimator}&\multicolumn{4}{|c|}{MLE}\\\cline{3-10}
&&Mean&Variance&(LCL, UCL)&CP&Mean&MSE&LCL, UCL&CP\\\hline
0.4&0.000104&0.000123&$1.524\times 10^{-7}$&$(0.000642,~0.000888)$&0.9698&0.000309&$4.746\times 10^{-7}$&$(0.001041,~0.001659)$&0.9862\\\hline
0.5&0.000977&0.001016&$4.575\times 10^{-6}$&$(0.003176,~0.005208)$&0.9636&0.001860&$8.889\times 10^{-6}$&$(0.003983,~0.007704)$&0.9830\\\hline
0.6&0.006046&0.005932&$5.819\times 10^{-5}$&$(0.009019,~0.020883)$&0.9450&0.008182&$7.887\times 10^{-5}$&$(0.009225,~0.025589)$&0.9701\\\hline
0.7&0.028247&0.028009&0.000543&$(0.017681,~0.073699)$&0.9500&0.031625&0.000556&$(0.014591,~0.077841)$&0.9559\\\hline
0.8&0.107374&0.107834&0.002879&$(0.002658,~0.213009)$&0.9599&0.108912&0.002606&$(0.008854,~0.208971)$&0.9523\\\hline
0.9&0.348678&0.349326&0.006850&$(0.187109,~0.511543)$&0.9520&0.341149&0.006688&$(0.180857,~0.501442)$&0.9487\\\hline
0.93&0.483837&0.483837&0.006413&$(0.326884,~0.640790)$&0.9518&0.476420&0.006768&$(0.315177,~0.636989)$&0.9569\\\hline
0.96&0.664832&0.665289&0.004414&$(0.535067,~0.795511)$&0.9490&0.668659&0.005214&$(0.527134,~0.810185)$&0.9626\\\hline
0.99&0.904382&0.904440&0.002129&$(0.813998,~0.994883)$&0.8560&0.935051&0.002511&$(0.836834,~1.000000)$&0.9739\\
\hline
\end{tabular}}
\end{center}
\end{landscape}
\pagebreak
\begin{center}{\bf Table 10: Efficiency of $\hat R_M$ with respect to $\hat R_M^{*}$}\\ $~r_{1}=10,~r_{2}=5,~\theta_{1}=0.7,~\theta_{2}=0.8,~R=0.8543644$\\
{\footnotesize\begin{tabular}{|c|c|c|c|c|}
\hline{$c_1|c_2$} &15&20&25&30\\
\hline10&170.3204&168.5565&162.8126&184.5733\\
\hline15&123.3550&119.4340&121.8824&124.4574\\
\hline20&107.4592&108.5157&110.6393&108.1562\\
\hline25&106.2693&104.6287&105.2644&103.2766\\\hline
\end{tabular}}
\end{center}
\begin{center}{\bf Table 11: Efficiency of $\hat R_M$ with respect to $\hat R_M^{*}$}\\ $~r_{1}=5,~r_{2}=10,~\theta_{1}=0.8,~\theta_{2}=0.7,~R=0.8212655$\\
{\footnotesize\begin{tabular}{|c|c|c|c|c|}
\hline{$c_1|c_2$} &15&20&25&30\\
\hline10&147.774&154.9791&157.9302&162.8920\\
\hline15&114.8463&118.9883&118.0901&121.116\\
\hline20&110.571&109.2363&106.1009&111.0485\\
\hline25&103.5806&104.4578&100.1903&103.9806\\\hline
\end{tabular}}
\end{center}
\begin{center}{\bf Table 12: Efficiency of $\hat R_M$ with respect to $\hat R_M^{*}$}\\ $~r_{1}=10,~r_{2}=5,~\theta_{1}=0.7,~\theta_{2}=0.8,~R=0.2234182$\\
{\footnotesize\begin{tabular}{|c|c|c|c|c|}
\hline{$c_1|c_2$} &10&15&20&25\\
\hline15&147.8276&113.0031&108.1288&103.5886\\
\hline20&159.4114&115.1075&105.5036&104.5615\\
\hline25&150.5815&117.2885&104.9115&102.4807\\
\hline30&140.7334&117.5558&107.1203&102.5214\\\hline
\end{tabular}}
\end{center}
\begin{center}{\bf Table 13: Efficiency of $\hat R_M$ with respect to $\hat R_M^{*}$}\\ $r_{1}=10,~r_{2}=5,~\theta_{1}=0.8,~\theta_{2}=0.7,~R=0.07639545$\\
{\footnotesize\begin{tabular}{|c|c|c|c|c|}
\hline{$c_1|c_2$} &10&15&20&25\\
\hline15&133.3005&109.2876&104.7854&103.7044\\
\hline20&140.5820&108.7528&101.8761&100.2199\\
\hline25&144.2547&107.1663&100.6334&101.4030\\
\hline30&146.9855&114.6641&102.8547&101.3404\\\hline
\end{tabular}}
\end{center}
\pagebreak
\begin{center}{\bf Table 14: MSEs of Estimator $R$}\\
\vspace{0.2in}
\begin{tabular}{|c|c|c|c|c|}
\hline{$r_2|r_1$} &5&10&15&20\\
\hline & 0.909091&9.09091$\times 10^{-6}$&9.09091$\times 10^{-11}$&9.09091$\times 10^{-16}$\\
&0.920557&0.000162&1.37222$\times 10^{-7}$&1.55348$\times 10^{-9}$\\
5&0.914516&0.343349&0.362465&0.344545\\
&0.005747&4.09776$\times 10^{-7}$&8.34178$\times 10^{-11}$&3.25312$\times 10^{-16}$\\
&0.006513&0.280415&0.300813&0.284422\\\hline
 &0.999999&0.909091&9.09091$\times 10^{-6}$&9.09091$\times 10^{-11}$\\
 &0.9999468&0.918347&0.000160&2.19983$\times 10^{-7}$\\
 10&0.999996&0.91221&0.350020&0.347819\\
 &4.47059$\times 10^{-08}$&0.005862&8.72553$\times 10^{-7}$&1.81692$\times 10^{-12}$\\
 &6.68094$\times 10^{-11}$&0.006658&0.288780&0.283411\\\hline 
 &1&.999999&.909091&9.09091$\times 10^{6}$\\
 &1&0.999964&0.916321&0.000191\\
 15&1&0.999998&0.909904&0.331967\\
 &2.63535$\times 10^{-13}$&2.19439$\times 10^{-8}$&0.005932&4.68203$\times 10^{-7}$\\
 &8.26455$\times 10^{-23}$&2.72526$\times 10^{-11}$&0.006783&0.271694\\\hline 
 &1&1&0.999999&0.909091\\
 &1&0.999999&0.999964&.914222\\
 20&1&1&0.999998&0.907743\\ 
 &1.05866$\times 10^{-18}$&9.20165$\times 10^{-13}$&2.28706$\times 10^{-8}$&0.006325\\ 
 &1.77863$\times 10^{-32}$&8.26455$\times 10^{-23}$&2.72526$\times 10^{-11}$&0.007260\\\hline
\end{tabular}
\end{center}
\pagebreak
\begin{center}{\bf Table 15: MSEs of Estimator $R$}\\
\vspace{0.2in}
\begin{tabular}{|c|c|c|c|c|}
\hline{$r_2|r_1$} &5&10&15&20\\
\hline &0.666667&0.020833&0.000651&2.03450$\times 10^{-5}$\\
&0.679177&0.026838&0.002103&0.000206\\
5&0.671748&0.021386&0.001522&0.001859\\
&0.012088&0.000737&2.33284$\times 10^{-5}$&5.11005$\times 10^{-7}$\\
&0.012973&0.000982&3.90682$\times 10^{-5}$&0.000701\\
\hline 
&0.989583&0.666667&0.020833&0.000651\\
&0.984516&0.668551&0.026998&0.001960\\
10&0.989622&0.660862&0.021230&0.001143\\
&0.000364&0.011225&0.000801&1.61857$\times 10^{-5}$\\
&0.002692&0.012219&0.000720&2.13649$\times 10^{-5}$\\
\hline 
&0.999674&0.989583&0.666667&0.020833\\
&0.998673&0.984826&0.669140&0.0265948\\
15&0.999563&0.990105&0.661459&0.020817\\
&1.50795$\times 10^{-5}$&0.000312&0.012265&0.000715\\
&6.954222$\times 10^{-6}$&0.000227&0.013318&0.000845\\
\hline 
&0.999989&0.999674&0.989583&0.666667\\
&0.999867&0.998716&0.984689&0.670912\\
20&0.999988&0.999662&0.989835&0.663109\\
&2.96125$\times 10^{-7}$&8.48119$\times 10^{-7}$&0.000345&0.012157\\
&2.92646$\times 10^{-8}$&1.94865$\times 10^{-6}$&0.000254&0.013208\\
\hline
\end{tabular}
\end{center}
\pagebreak
\begin{center}{\bf Table 16: MSEs of Estimator $R$}\\
\vspace{0.2in}
\begin{tabular}{|c|c|c|c|c|}
\hline{$r_2|r_1$} &5&10&15&20\\
\hline &0.238095&7.61904$\times 10^{-5}$&2.43809$\times 10^{-8}$&7.80190$\times 10^{-12}$\\
 &0.239828&0.000384&3.17275$\times 10^{-6}$&7.44942$\times 10^{-8}$\\
 5&0.242251&0.028820&0.031522&0.029636\\
 &0.009815&1.03719$\times 10^{-6}$&4.49468$\times 10^{-10}$&6.71449$\times 10^{-13}$\\
 &0.007733&0.005539&0.006393&0.005711\\
 \hline 
 &0.750339&0.238095&7.61905$\times 10^{-5}$&2.43809$\times 10^{-8}$\\
 &0.753009&0.236090&0.003744&3.40036$\times 10^{-6}$\\
 10&0.742105&0.239565&0.033643&0.032150\\
 &0.010630&0.008706&9.65263$\times 10^{-7}$&4.16401$\times 10^{-10}$\\
 &0.012344&0.006546&0.006595&0.006113\\
 \hline 
 &0.918191&0.750339&0.238095&7.61905$\times 10^{-5}$\\
 &0.918237&0.756132&0.240675&0.000352\\
 15&0.920495&0.746061&0.243698&0.032421\\
 &0.003296&0.012589&0.009720&8.85057$\times 10^{-7}$\\
 &0.004096&0.014265&0.007625&0.006236\\
 \hline 
 &0.973193&0.918191&0.750339&0.238095\\
 &0.968532&0.919645&0.767043&0.240174\\
 20&0.973493&0.922001&0.757469&0.242044\\
 &0.000989&0.003383&0.011126&0.009538\\
 &0.001103&0.004192&0.012577&0.007625\\\hline
\end{tabular}
\end{center}
\pagebreak
\begin{center}{\bf Table 17: MSEs of Estimator $R$}\\
\vspace{0.2in}
\begin{tabular}{|c|c|c|c|c|}
\hline{$r_2|r_1$} &5&10&15&20\\
\hline &0.526316&0.310784&0.183515&0.108364\\
&0.526810&0.299460&0.174113&0.103238\\
5&0.526508&0.308696&0.182257&0.105726\\
&0.017799&0.012224&0.007922&0.0043331\\
&0.015585&0.011825&0.008655&0.005117\\
\hline 
&0.720294&0.526316&0.310784&0.183515\\
&0.726634&0.529795&0.305464&0.175418\\
10&0.716990&0.529132&0.314470&0.183407\\
&0.011833&0.016580&0.013347&0.008355\\
&0.011819&0.014537&0.012886&0.009171\\
\hline 
&0.834836&0.720294&0.526316&0.313784\\
&0.840993&0.722299&0.530707&0.305661\\
15&0.833705&0.712836&0.529979&0.314475\\
&0.007389&0.012591&0.015473&0.013614\\
&0.008308&0.012569&0.013530&0.013235\\
\hline 
&0.902473&0.834836&0.720294&0.526316\\
&0.903279&0.840593&0.726930&0.525431\\
20&0.900223&0.833145&0.717270&0.525064\\
&0.003855&0.006583&0.012184&0.017255\\
&0.004660&0.007377&0.012206&0.015104\\\hline
\end{tabular}
\end{center}
\pagebreak
\begin{center}{\bf Table 18: MSEs of Estimator $R$}\\
\vspace{0.2in}
\begin{tabular}{|c|c|c|c|c|}
\hline{$r_2|r_1$} &5&10&15&20\\
\hline&0.952381&0.321076&0.102261&0.033509\\
&0.954791&0.300429&0.107198&0.038337\\
5&0.953994&0.313513&0.107466&0.033587\\
&0.001314&0.012567&0.005071&0.001269\\
&0.001318&0.013776&0.006272&0.001443\\
\hline
&0.999985&0.952381&0.312076&0.102261\\
&0.999870&0.952055&0.296181&0.103630\\
10&0.999989&0.951173&0.309150&0.103415\\
&1.49867$\times 10^{-7}$&0.001352&0.012463&0.004663\\
&6.16543$\times 10^{-9}$&0.001373&0.013512&0.005783\\
\hline
&1&.999985&.952381&.312076\\
&0.999998&0.999865&0.952267&0.302442\\
15&1&0.999983&0.951499&0.315536\\
&1.01544$\times 10^{-10}$&2.91485$\times 10^{-7}$&0.001478&0.012921\\
&2.37772$\times 10^{-17}$&4.87421$\times 10^{-8}$&0.001480&0.014082\\
\hline
&1&1&0.999985&0.952381\\
&1&.999998&0.999858&0.952217\\
20&1&1&0.999986&0.951531\\
&4.40539$\times 10^{-13}$&1.37763$\times 10^{-10}$&1.86115$\times 10^{-7}$&0.001436\\
&2.43492$\times 10^{-24}$&2.37772$\times 10^{-17}$&8.33376$\times 10^{-9}$&0.001436\\\hline
\end{tabular}
\end{center}
\pagebreak
\begin{landscape}
\pagebreak
\begin{center}{\bf Table 19: Calculation of Confidence Interval and CP of $R$}\\$n_1=n_2=10,~\theta_1=\theta_2=0.9,~Repetation=10000$\\
\vspace{0.2in}
{\footnotesize\begin{tabular}{|c|c|c|c|c|c|c|c|c|c|}
\hline{$(r_1,~r_2)$}&Reliability &\multicolumn{4}{|c|}{Unbiased Estimator}&\multicolumn{4}{|c|}{MLE}\\\cline{3-10}
&&Mean&Variance&(LCL, UCL)&CP&Mean&MSE&LCL, UCL&CP\\\hline
$(5,~20)$&0.108363&0.108382&0.005508&$(0.037085,~0.253851)$&0.9540&0.104635&0.004648&$(0.028995,~0.238266)$&0.9372\\\hline
$(5,~15)$&0.183515&0.184696&0.009011&$(0.001353,~0.370744)$&0.9650&0.176514&0.008193&$(0.000888,~0.353917)$&0.9510\\\hline
$(5,~10)$&0.310784&0.309366&0.012371&$(0.091370,~0.527362)$&0.9540&0.300321&0.012822&$(0.078386,~0.522256)$&0.9592\\\hline
$(5,~5)$&0.526316&0.524415&0.014289&$(0.290129,~0.758701)$&0.9504&0.524652&0.016296&$(0.274403,~0.774854)$&0.9627\\\hline
$(10,~5)$&0.720294&0.721299&0.011505&$(0.511075,~0.931525)$&0.9545&0.730976&0.011572&$(0.520138,~0.941814)$&0.9560\\\hline
$(15,~5)$&0.834836&0.832709&0.008217&$(0.672409,~1.000000)$&0.9629&0.840145&0.007324&$(0.672409,~1.000000)$&0.9470\\\hline
$(20,~5)$&0.902472&0.902443&0.004755&$(0.767295,~1.000000)$&0.9545&0.905403&0.003967&$(0.781953,~1.028853)$&0.9545\\\hline
\end{tabular}}
\end{center}
\end{landscape}
\end{document}